\documentclass[prd,aps,amsmath,amssymb]{revtex4}

\usepackage{graphicx}

\begin{document}
\title{
Partial Deconfinement in Color Superconductivity }
\author{Francesco {\sc Sannino}$^1$ \quad Nils {\sc
Marchal}$^{1,2}$ \quad Wolfgang {\sc Sch\"afer}$^1$ }
\address{$^1$ NORDITA, Blegdamsvej 17, DK-2100 Copenhagen \O, Denmark \\
$^2$ LAPTH, F-74941 Annecy-le-Vieux Cedex, France }
\date{February 2001}

\begin{abstract}
 We analyze the fate of the unbroken SU(2) color gauge interactions for 2
light flavors color superconductivity at non zero temperature.
Using a simple glueball lagrangian model we compute the
deconfining/confining critical temperature and show that is
smaller than the critical temperature for the onset of the
superconductive state itself. The breaking of Lorentz invariance,
induced already at zero temperature by the quark chemical
potential, is shown to heavily affect the value of the critical
temperature and all of the relevant features related to the
deconfining transition. Modifying the Polyakov loop model to
describe the SU(2) immersed in the diquark medium we argue that
the deconfinement transition is second order. Having constructed
part of the equation of state for the 2 color superconducting
phase at low temperatures our results may be relevant for the
physics of compact objects featuring a two flavor color
superconductive state.
\end{abstract}

\maketitle

\section{Introduction}
\label{uno}

Quark matter at very high density is expected to behave as a color
superconductor \cite{REV}. Possible physical applications are
related to the physics of compact objects \cite{REV}, supernovae
cooling \cite{Carter:2000xf} and explosions \cite{HHS} as well as
to the Gamma Ray Bursts puzzle \cite{OS}. Here we concentrate on
some features related to color superconductivity with 2 light
flavors (2SC). The low-energy effective Lagrangian describing the
in medium fermions and the broken sector of the $SU_c(3)$ color
groups for the 2 flavor color superconductor (2SC) has been
constructed in Ref.~\cite{CDS,OS2}. The 3 flavor case (CFL) has
been developed in \cite{CG,{Casalbuoni:1999zi}}. The effective
theories describing the electroweak interactions for the
low-energy excitations in the 2SC and CFL case can be found in
\cite{CDS2001}. The global anomalies matching conditions and
constrains are discussed in \cite{S}. An interesting property of
the $2SC$ state is that the three color gauge group breaks
spontaneously to a left over $SU(2)$ subgroup and it can play a
role for the physics of compact objects \cite{OS}. In Reference
\cite{rischke2k} it has been shown that the confining scale of the
unbroken $SU(2)$ color subgroup is lighter than the
superconductive gap $\Delta$. The confined degrees of freedom,
glueball-like particles, are expected to  be light with respect to
$\Delta$, and the effective theory based on the anomalous
variation of the dilation current has been constructed in
\cite{OS2}.

Clearly for the physics of compact objects and more generally for
a complete understanding of the QCD phase diagram it is relevant
to know at what temperature the $SU(2)$ color gauge group
confines/deconfines, the order of the phase transition and the
equation of state.

Investigating the deconfinement phase transition is, in general, a
complex problem. At zero density importance sampling lattice
simulations are able to provide vital information about the nature
of the temperature driven phase transition for 2 and 3 colors
Yang-Mills theories with and without matter fields (see
\cite{Karsch:2001jb} for a review). Different models
\cite{Campbell:ak,Simonov:bc,Agasian:fn,Sollfrank:du,Carter:1998ti,Schaefer:2001cn,Drago:2001gd,Renk:2002md,{Pisarski:2001pe},DP,{KorthalsAltes:1999cp},{Dumitru:2001xa},Scavenius:2002ru,{Wirstam:2001ka}}
are used in literature to tackle/study the features of this phase
transition from a theoretical stand point. Some models compute non
zero temperature corrections for the glueball Lagrangian with or
without elementary gluon degrees of freedom (the latter added to
describe the deconfined side of the phase transition). Others rely
on mean field theories encoding the symmetries of the Polyakov
loops \cite{Pisarski:2001pe}.

Hence we consider a simple, but predictive, model for the
deconfinement temperature which makes use of the glueball
Lagrangian valid at non zero quark density \cite{OS2}. We
investigate the one loop thermal effective potential corrections
for the dilatonic Lagrangian and observe that as we increase the
temperature a new local minimum sets in for a lower value of the
gluon condensate with respect to the zero temperature one. The
critical temperature is defined as the value for which the two
local minima have the same free energy value. Above this critical
temperature the model is no longer applicable since new degrees of
freedom like the unconfined gluons are expected to appear (see
e.g. \cite{Carter:1998ti}) and we will briefly comment on their
effects. An amusing feature of the model is that the critical
temperature can be determined analytically. This is so since the
new minimum appears for a zero vacuum expectation value of the
gluon condensate \footnote{In the full theory it is more
reasonable to expect just a drastic drop of the condensate close
to the phase transition.} and at this point, in the one loop
approximation, one can exactly compute the effective thermal
potential yielding the following estimate for the critical
temperature:
\begin{equation}
T_{c}=\sqrt[4]{\frac{90\,v^3}{2e\pi^2}}\hat{\Lambda} \ .
\label{ApproxTc}
\end{equation}
Here $e$ is the Euler number, $\hat{\Lambda}$ is the confining
scale of the $SU(2)$ gluon-dynamics in 2SC and $v$ is the gluon
\cite{rischke2k} as well as light glueball velocity \cite{OS2}.
Equation (\ref{ApproxTc}) is a good approximation also for the in
vacuum theory. Here when Eq.~(\ref{ApproxTc}) is adjusted to take
into account the gluonic degrees of freedom the higher order (in
loop) contributions are shown to be less than 10\% (see
\cite{Carter:1998ti}).

We find that the deconfining/confining critical temperature is
smaller than the critical temperature $T_{2SC}$ for the
superconductive state itself which is estimated to be $T_{2SC}
\approx 0.57~\Delta$ with $\Delta$ the 2SC gap \cite{PR}. Actually
the breaking of Lorentz invariance, due to the quark chemical
potential and encoded in the glueball velocity, further reduces
the critical temperature by a factor $v^{3/4}$ relative to the in
vacuum case. This is a general feature independent of the model
Lagrangian, also observed in \cite{Casalbuoni:1999zi}. The
temperatures in play are much less than the value of the quark
chemical potential. {}The situation is more involved if also
rotational invariance breaks spontaneously due, for example, to
the appearance of a spin one condensate \cite{Sannino:2001fd}. We
study the glueball mass as function of temperature, chemical
potential and $\Delta$. In the confined phase the mass is, at a
very good approximation, constant with respect to the temperature.

By computing the glueball thermal effective potential we provide
part of the equation of state for the 2SC phase at low
temperatures. In particular we can compute the pressure, the
energy density and the entropy of the system below the critical
deconfining temperature.

It is important to stress that in this paper we are considering an
ideal 2SC state where the up and down quarks are massless and the
strange quark is infinitely massive. When computing properties
related to the physics of compact stars it is very important to
introduce in the model the effects of the quark masses as well as
the ones induced by a not too heavy strange quark. These effects
may affect the $SU(2)$ gluon properties and can be investigated
using for example the effective theories near the fermi surface
\cite{REV}. It would also be very interesting to see how the non
perturbative $SU(2)$ dynamics might affect the recent results in
\cite{Alford:2002kj}. However even within the present restrictive
framework our estimate for the $SU(2)$ confining temperature may
be a useful guide for astrophysical models of compact stars like
the one in Ref.~\cite{OS} featuring a 2SC state.

Since the gluon-condensate is not a true order parameter for the
deconfining transition the glueball Lagrangian cannot be used
 to infer the order of the transition itself.
 To settle this issue we modify the Polyakov's loop  inspired model \cite{Pisarski:2001pe}
to fit the present case and finally predict a second order phase
transition. Finally we suggest how ordinary lattice importance
sampling techniques can be used to check our results and
constitute, at the same time, the first simulations testing the
high quark chemical potential but small temperature region of the
QCD phase diagram.

The glueball Lagrangian, if extended to describe the transition
point, predicts a first order transition which seems to disagree
with the prediction based on the order parameter (Polyakov's
loop). Actually the disagreement is an apparent one. Indeed only
the order parameter is obliged to know about the order of the
phase transition. Any other gauge invariant quantity does not need
to display the same behavior while still bearing information about
the phase transition itself (see for example
\cite{Pisarski:2001pe} page 3 equations (9) and (10) and
subsequent discussion (first reference)). In practice the critical
temperature predicted represents also the limit of applicability
of our simple model. At the critical point the Ginzburg-Landau
theory for the order parameter is the proper way to describe the
transition itself and can be used to infer the order of the phase
transition. Unfortunately though the Ginzburg-Landau theory cannot
predict the critical temperature. The previous discussion does not
imply that the glueball and the order parameter (the Polyakov
loop) at the transition  are not related \cite{Sannino:2002wb}.

It is important to stress that in general we have a tower of
scalar, pseudoscalar and other excited glueball states in the
confined regime together with the other physical states involving
quarks of the 2SC state. We have made the standard assumption that
the low energy $SU(2)$ dynamics is dominated by the associated
lightest mode in the theory: the scalar glueball. This state does
not couple to the light ungapped up and down quarks in the
direction 3 of color (for a review of the complete low energy
effective theory of the 2SC state see the 7th reference in
\cite{REV}). Besides, according to \cite{Litim:2001je}, the quark
temperature effects are exponentially suppressed ($\sim
\exp({-\Delta/T})$) so for $T<T_c$ and $\hat{\Lambda}<\Delta$, for
an initial investigation, we can neglect these corrections. For
temperatures in the range $T_c < T < T_{2SC}$ the gapped quark
dynamics is no longer negligible and some of their effects have
been computed using transport theory  in \cite{Litim:2001je}. Our
model must be considered only as a first step toward a more
complete theory of the 2SC state where the $SU(2)$ non
perturbative dynamics is included.

In Section \ref{Glueball} we provide the light glueball Lagrangian
and construct the one loop thermal effective action. Here we
suggest a way to relate the results obtained by employing
different parameterizations for the glueball field. In Section
\ref{Features} we study the relevant features connected with the
deconfining transition. We provide an economical criterion to
estimate the critical temperature similar to the one extensively
used in literature for the in vacuum Yang-Mills theories
\cite{Carter:1998ti}. Finally using the Polyakov loop model
adapted to the present case we show the phase transition to be
likely second order. We conclude in Section \ref{Conclusions}.

\section{Glueball Effective Lagrangian at finite Temperature}
\label{Glueball}

The light glueball action for the in--medium Yang-Mills theory is
\cite{OS2}:
\begin{eqnarray}
S_{Glueball}=\int
&d^4x&\left\{\frac{c}{2}\,H^{-\frac{3}{2}}\left[\partial^{0} H
\partial^{0}H - v^2
\partial^iH
\partial^iH\right] -\frac{1}{2}H\log\left[\frac{H}{\hat{\Lambda}^4}\right] \right\} \ .
\label{G-ball}
\end{eqnarray}
$H$ is the composite field describing, upon quantization, the
scalar glueball
\cite{{OS2},schechter,{SS},{SSSusy},{MS},{SST},{GJJK},{CE}} in
medium and possesses mass-scale dimensions four. Here $c$ is a
positive constant \footnote{Here we absorbed the coefficient $b$
present in the Lagrangian of \cite{OS2} in the definition for $H$.
This coefficient is relevant when comparing the results of the
glueball Lagrangian derived for different number of colors and
flavors \cite{SS}.} which fixes the tree glueball mass. Our
results do not depend on the specific value assumed by this
constant. It is also important to stress that the glueballs move
with the same velocity as the underlying gluons in the 2SC color
superconductor \cite{OS2}. The velocity depends on the gluon
dielectric constant ($\epsilon$) and magnetic permeability
($\lambda$) via $v=1/\sqrt{\epsilon \lambda}$. The dielectric
constant $\epsilon$ is different from unity (in fact $\epsilon \gg
1$  in the 2SC case \cite{rischke2k}) leading to an effectively
reduced gauge coupling constant. Studying the polarization tensor
at asymptotically high quark densities for the $SU(2)$ gluons in
\cite{rischke2k} was found:
\begin{equation}
\epsilon =1 + \frac{g_s^2 \mu^2}{18 \pi^2 \Delta^2}\ , \qquad
\lambda =1 \ , \label{el}
\end{equation}
with $g_s$ the underlying $SU(3)$ coupling constant and $\mu$ the
quark chemical potential. This result has also been derived via
effective theories valid close to the Fermi surface
\cite{Casalbuoni2001}. In the effective Lagrangian $\hat{\Lambda}$
is a physical constant related to the confining scale of the in--
medium 2 color Yang-Mills theory. Following \cite{rischke2k} we
have the one loop relation:
\begin{eqnarray}
\hat{\Lambda}=\Delta \exp \left[-\frac{8\pi^2}{b g_s^2(\mu)}{
\sqrt{\frac{\epsilon(\mu/\Delta)}{\lambda(\mu/\Delta)}}}\right]\simeq
\Delta \exp \left[-\frac{2\sqrt{2}\pi}{11}
\frac{\mu}{g_s(\mu)\Delta} \right]\ , \label{lambda}
\end{eqnarray}
with $b=22/3$ (at one loop) for $SU(2)$ and in the last step we
considered the asymptotic solution of Ref.~\cite{rischke2k}, for
convenience reported in Eq.~(\ref{el}). By using
$\Lambda_{QCD}\simeq 300$ MeV, $\mu \simeq 500$ MeV and a gap
value of about $30$~MeV one gets $\hat{\Lambda} \simeq 1$~MeV. It
is hence reasonable to expect that the glueballs are light (with
respect to the gap) with a mass typically somewhat larger or of
the order of the confining scale. They are stable with respect to
the strong interactions unlike ordinary glueballs while still
decaying into two photons \cite{OS2}. The potential in
Eq.~(\ref{G-ball}) can be considered a zeroth order model
\cite{schechter,{MS},SST} for a Yang-Mills theory in medium
\cite{OS2} in which the glueballs are the associated hadronic
particles. The minimum of the potential $V$ (see \cite{OS2} for
details) is taken for
\begin{equation}
\langle H \rangle =\frac{\hat{\Lambda}^4}{e} \ , \quad {\rm
at~which~point} \quad
\langle{V}\rangle=-\frac{\hat{\Lambda}^4}{2\,e} \ .
\end{equation}
For the zero density case a number of phenomenological questions
have been discussed using this type of toy model Lagrangian
Eq.~(\ref{G-ball}) \cite{SST}.

In order to extract dynamical information we define a canonically
normalized (with canonical mass dimension one) glueball field $h$
via:
\begin{equation}
H=f(h)=f_{(0)}^4+ f_{(1)} h+f_{(2)}\frac{h^2}{2!} +\cdots +
f_{(n)}\frac{h^n}{n!}+\cdots \ ,
\end{equation}
where we require $f(h)$ to be a well behaved function of the
glueball field $h$ with non vanishing $f_{(0)}$ and $f_{(1)}$. The
normalization condition of the kinetic term, at the tree level,
yields the constraint:
\begin{equation}
c^{\frac{1}{2}}\, f_{(1)}= f_{(0)}^{3} \ .
\end{equation}

It is reasonable to expect that any interpolating function $f(h)$
should lead to the same physical results. This is indeed the case
at the tree level since all of the possible choices to define a
canonically normalized field are equivalent.

However when considering thermal/quantum corrections is hard to
demonstrate that different choices lead to the same physical
results. We remind the reader of the time-honored sigma model
example where the linear version is a renormalizable theory while
the non linear sigma model is {\it not} a renormalizable theory in
the usual sense.

\begin{figure}[hbtp]
\begin{center}
\includegraphics[angle=0]{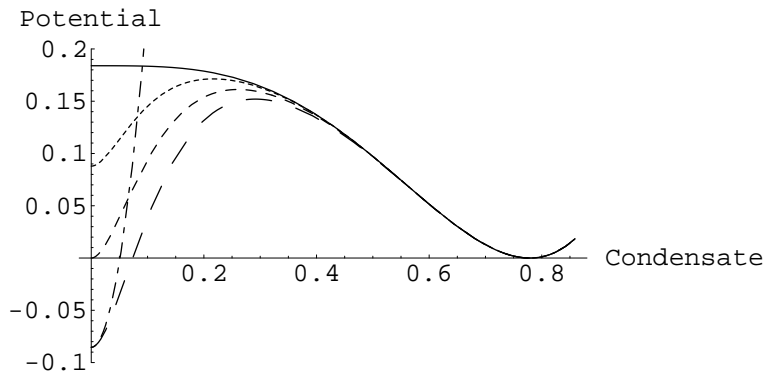}
\end {center}
\caption{Potential function
$V\left[\bar{h}\right]/\hat{\Lambda}^4$ for $\mu=500$~MeV and
$\Delta=30$~MeV as function of the condensate
$\bar{h}/\hat{\Lambda}$ for different values of the temperature.
The solid line corresponds to $T=0$; the dotted line to
$T=0.85~T_c$; the short-dashed to $T=T_c$; the long dashed to
$T=1.1~T_c$. Finally the dot-dashed line corresponds to the high
temperature expansion near $\bar{h}=0$ for $T=1.1~T_c$. We have
chosen for definitiveness $c=1/(50 \sqrt{e})$, corresponding to a
zero temperature glueball mass of $5\hat{\Lambda}$.}
\label{figure1}
\end{figure}

In order to keep our results as independent as possible from the
specific function $f(h)$ here we define thermal averages directly
in terms of $H$. More specifically following Dolan and Jackiw
\cite{Dolan:qd} we formally introduce the temperature effective
action $\Gamma(\overline{H})$ -the generating functional for
single-particle irreducible Green's functions via:
\begin{eqnarray}
W[J]&=&-i\log\left[\frac{{\rm Tr}e^{-\frac{\cal H}{T}}\exp\left[i
\int d^4x \, H(x)J(x) \right]}{{\rm Tr}e^{-\frac{\cal H}{T}}
}\right] \ , \\  \overline{H}(x)&=&
 \frac{\delta W[J]}{\delta
J(x)}\ , \label{classyfield}
\\
\Gamma[\overline{H}]&=& W[J]-\int d^4x\, {\overline
 H}(x) J(x) \ . \label{classyaction}
\end{eqnarray}
${\cal H}$ is the Hamiltonian and $J(x)$ the external source for
the gluon condensate.  In the last equation $J(x)$ is eliminated
in favor of $\overline{H}(x)$ by the definition in
(\ref{classyfield}). We also have that ${\delta
\Gamma[\overline{H}]}/{\delta \overline{H}(x)=-J(x)}$ and
$\overline{H}(x)$, evaluated at $J=0$, is the thermodynamic
average of the gluon condensate field $H(x)$. The present
definition of the effective action is independent of the choice of
the interpolating field function. For the present purposes it is
sufficient to study $\Gamma[\overline{H}]$ for constant
${\overline H}(x)$ and consider the effective potential:
\begin{equation}
V[\overline{H}]=-\frac{\Gamma[\overline{H}]}{\rm{space-time~volume}}
\ .
\end{equation}
In practice the $J$ dependent tree generating functional for the
trace anomaly is (with $V\left[J\right] = -
W\left[J\right]/(\rm{space-time~volume})$ for constant fields)
\begin{equation}
V_{Tree}[J]=\frac{1}{2}
H\log\left[\frac{H}{\hat{\Lambda}^4}\right] -J\, H \ .
\end{equation}
and the one loop thermal effective potential as function of $J$
is:
\begin{eqnarray}
V[J]&=&{2} f_{(0)}^4\log\left[\frac{f_{(0)}}{\hat{\Lambda}}\right]
- J\, f_{(0)}^4 + \frac{T}{v^3 \, 2\pi^2} \int_0^\infty dk\,k^2
\log \left[1 - \exp \left({-\frac{\epsilon_J}{T}}\right)\right] \
,
\end{eqnarray}
where  $\displaystyle{\epsilon_J= \sqrt{k^2+ M^2_J \left(
f_{(0)},f_{(2)} \right)}}$, and $M^2_J$ is defined via the
curvature of the potential as
\begin{equation} M^2_{J}= \left. \frac{\partial^2 V}{\partial h^2}
\right|_{h=0}= \frac{f_{(0)}^2}{2c}
+f_{(2)}\left[2\log\frac{\sqrt[4]{e}\, f_{0}}{\hat{\Lambda}
}-J\right] \ . \label{curvatura}
\end{equation}
With the help of
\begin{equation}
\overline{H} \equiv \overline{h}^4=-\frac{\delta V\left[J\right]}{
\delta J} = f_{(0)}^4 + \frac{f_{(2)}}{v^3 4\pi^2} \,
\int_0^\infty  \frac{dk \, k^2}{\left[
\exp\left(\displaystyle{\frac{\epsilon_J}{T}}\right) -1\right]\,
\epsilon_J} \ ,
\end{equation}
we deduce the effective potential
\begin{equation}
V[\overline{h}]=\left[1 - J\,\frac{\partial}{\partial
J}\right]V[J] \ ,
\label{finalV}
\end{equation}
where the functional derivative with respect to $J$ is replaced
with a partial derivative since we are now dealing with constant
fields. We need  now to solve for $J[\overline{H}]$  as function
of $\overline{H}$ and then extremize the action. We identify
$\bar{H}$ with $\bar{h}^4$ only after $J$ has been eliminated. For
a general choice the function $f(h)$ one cannot find an analytical
expression for $J[\overline{H}]$. However we immediately notice
that for $f_{(2)}=0$ (at the one-loop level) there is no
dependence on $J$ and we have $\overline{h}=f_{(0)}$ as well as a
positive definite curvature $M^2={f_{(0)}}^2/{2c}$. To be more
specific our glueball field function is now $f(h)=f_{(0)}^4 +
f_{(1)}\,h$ where we truncate our function to the quadratic term
since higher terms do not affect the one loop result. Actually any
function $f(h)$ with just vanishing $f_{(2)}$ leads to the same
source independent effective thermal potential:
\begin{eqnarray}
V\left[\bar{h}\right]= {\hat{\Lambda}^4 \over 2 e} +
2\bar{h}^4\log\left[\frac{\bar{h}}{\hat{\Lambda}}\right] +
\frac{T^4}{v^3 \, 2\pi^2} \int_0^\infty dx\,x^2 \log \left[1 -
\exp \left(-\sqrt{x^2 + \frac{\bar{h}^2}{2cT^2}} \right) \right] \
, \label{thermalP}
\end{eqnarray}
where for convenience we subtracted the constant value of the
potential evaluated on the vacuum at zero temperature. This
expression is well defined for any value of $\bar{h}$.
$V\left[\bar{h}\right]$ is shown in Fig. \ref{figure1} for
different values of the temperature and a given value of $c$ which
fixes the zero temperature tree-level glueball mass (i.e.
$M^2=\hat{\Lambda}^2/2\sqrt{e}c$). The plot is provided only for
illustration and the general feature of the potential does not
change for different choices of the chemical potential and
reasonable values of the gap parameter.  Note that our results for
the critical temperature (presented in the next section) are
evaluated at different values of the quark chemical potential and
 the gap $\Delta$.  As we increase the temperature we observe a new minimum
setting in for $\bar{h}$ at zero. We also note that the position
of the old minimum is not much affected by temperature corrections
over a large range of temperatures (see Fig.~\ref{figure1}). Close
to the new minimum at zero is possible to perform the high
temperature expansion leading to:
\begin{eqnarray}
\lim_{\bar{h}\rightarrow 0}V\left[\bar{h}\right]={\hat{\Lambda}^4
\over 2 e} - \frac{2\pi^2}{90}\frac{T^4}{v^3} +
\frac{\bar{h}^2}{2c}\frac{T^2}{24 v^3} + {\cal
O}\left(\bar{h}^4\right) \ . \label{exactTP}
\end{eqnarray}

Before describing in some detail the features of the phase
transition we now briefly comment on another possible choice of
the glueball field widely used in literature. This is the
exponential representation:
\begin{eqnarray}
H=f_{(0)}^4 \exp\left[\frac{h}{f_{(0)}\sqrt{c}}\right] \ .
\end{eqnarray}
This function recovers the previous one for small field
fluctuations. However since $f_{(2)}=f_{(0)}^2/c$ is not vanishing
we cannot deduce an analytical expression of $J$ as function of
$\bar{h}$. Note that if we would naively set $J$ to zero from the
beginning the second derivative of the potential defined in
Eq.~(\ref{curvatura}) is not definite positive for all values of
$\bar{h}$ and the integral in Eq.~(\ref{thermalP}) is ill defined.
Often
in the literature the thermal corrections are computed without
including the source $J$. We have shown that, at the one loop
level, the linearly realized representation is not affected by the
introduction of the source term while the non linear realization
used for example in \cite{Drago:2001gd} are very much affected and
should be handled with care. We expect the partial derivative term
$J\partial V[J]/\partial J$ in Eq.~(\ref{finalV}) to help
compensating for the possible different choices of $f(h)$. In the
rest of this work we shall use the linear realizations.

Clearly after having defined the extremum of the effective
potential it is a simple matter to derive all of the relevant
thermodynamical quantities. For the reader's convenience we
summarize the standard relations between the thermodynamical
quantities and the free energy (per unit volume) $F=V$ (V is
evaluated on the minimum) with the pressure $P=- F$ while the
entropy per unit volume is $S=-\partial F/\partial T$.

\section{Relevant features of the deconfining transition}
\label{Features} Studying the one loop thermal effective potential
in Eq.~(\ref{finalV}) one observes  that when increasing the
temperature a new local minimum sets in at $\bar{h}=0$ and, for a
certain range of temperatures, the potential has two local minima.
The temperature for which the two minima have the same free energy
is:
\begin{equation}
T_{c}=\sqrt[4]{\frac{90\,v^3}{2e\pi^2}}\hat{\Lambda} \ .
\end{equation}
This value is obtained by comparing the jump of the potential due
to the temperature corrections (actually at zero $\bar{h}$) with
the respect to the zero temperature minimum, and it does not
depend on the specific value assumed by the constant $c$ in the
effective Lagrangian. The latter can be fixed once the glueball
mass is known.

\begin{figure}[hbtp]
\begin{center}
\includegraphics[width=16cm, height=5cm, angle=0]{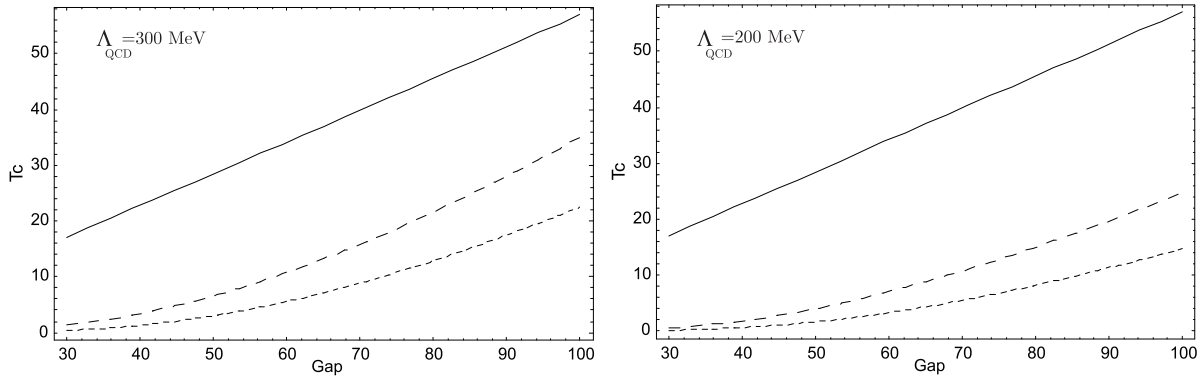}
\end {center}
\caption{Plots of the $SU(2)$ critical temperature for 2 values of
the quark chemical potential ($\mu=400$~MeV long--dashed line;
$\mu=500$~MeV short--dashed line) as function of the
superconductive gap $\Delta$. The solid line corresponds to the
critical temperature for the superconductive state $0.57 \Delta$.
The left panel corresponds to $\Lambda_{QCD}=300$~MeV while the
right one corresponds to $\Lambda_{QCD}=200$~MeV} \label{figure2}
\end{figure}

Assuming that the drop in the gluon condensate together with the
drastic change in the glueball mass are related to the
deconfinement phase transition as supported by lattice simulations
\cite{Bacilieri} we interpret Eq.~(\ref{ApproxTc}) as an estimate
for the critical temperature. In figure \ref{figure2} we plot the
critical temperature as function of the superconductive gap for
different values of the quark chemical potential. In models where
the contribution of the elementary gluons is added one finds a
smaller temperature (see for example \cite{Carter:1998ti}). The
reduction is due to the extra contribution of the {\it light}
gluons appearing at $T_c$. This effect can be estimated assuming
that the main contribution of the gluons at $T_c$ is the free
energy for unconfined gluons propagating with velocity $v$. By
simply adding to the effective thermal potential the term $-2\pi^2
\left[2(N^2-1)\right] T^4\Theta (T - T_c)/(90v^3)$ for a general
number of colors $N$ the temperature for which the two minima have
the same free energy value is lowered to
\begin{equation} {T}_{c} \rightarrow \frac{T_c}{\sqrt[4]{2\left(N^2-1\right)+1}} \ .
\label{secondTc}\end{equation} The reduction is perhaps too
drastic since, in many investigations at zero density, it has been
argued that a better fit to the Lattice data even at temperatures
as high as 3 times the critical temperature requires an effective
number of gluon degrees of freedom lower than the one predicted by
a free gas approximation. It is then quite likely that the true
critical temperature lies in between the one presented in
Eq.~(\ref{ApproxTc}) computed without gluons and the one estimated
Eq.~(\ref{secondTc}).

When reducing the temperature from the quark gluon plasma phase we
see that color superconductivity first sets in along the
temperature axis  with the $SU(2)$ of color still unconfined and
finally the $SU(2)$ confines at a lower value of the temperature
(see Fig.~\ref{figure2}). Higher order corrections to our critical
temperature are shown to be smaller than 10\% (see
\cite{Carter:1998ti}).

We also see confronting the left and right panel of
Fig.~\ref{figure2} that the critical temperature decreases if
$\Lambda_{QCD}$ decreases.

Our model applies directly only to the ideal 2SC state and for
physical applications we need to consider in some detail the
corrections induced for example by the quark masses. Nevertheless
it might be instructive to show how the explicit dependence of the
$SU(2)$ confining temperature on $\mu$ and $\Delta$ may be helpful
to astrophysical applications. Indeed in a model for Gamma Ray
Bursts (GRBs) \cite{OS} it was suggested that some compact stars
might feature a hot 2SC surface layer. The GRBs model used the
glueballs as an active degree of freedom. So we need to know when,
along the temperature axis, the 2SC layer enters the $SU(2)$
confining regime. Within our model calculations we indicate in the
$T-\mu$ phase diagram where the glueballs degrees of freedom start
playing a role. For example if $\mu=400-500$ MeV from
Fig.~\ref{figure2} we deduce that the $SU(2)$ confines at
$T_c\approx 10$~MeV provided $\Delta \geq 60-70$~MeV. Our work
might also be useful when investigating the cooling process in
compact stars.

The derived free energy for the $SU(2)$ glue for very low
temperatures represents an initial step when computing part of the
complete equation of state which is needed when considering the
thermodynamics of compact object featuring a 2SC state. In our
model we have assumed the glueball velocity not to depend on the
temperature. This is reasonable since the temperature corrections
for $v$ are exponentially suppressed, more specifically the
suppression factor is $e^{-\Delta/T}$
\cite{Rischke:2001py,{Litim:2001je}}. Since we find the critical
temperature to lie well below the critical temperature for color
superconductivity ($T_{2SC}\approx 0.57 \Delta$) our results
provide a consistent picture. It is important to stress that for
temperatures $T_c < T < T_{2SC}$ the gapped quark contributions
are no longer negligible. In Ref.~\cite{Litim:2001je} using the
transport theory some relevant temperature effects have been
analyzed.

It is useful to study the dependence of the glueball mass on the
temperature. Defining the square of the mass as the potential
curvature evaluated at the (global) minimum we observe (see
Fig.~\ref{figure1}) that for $T<T_c$ the curvature is practically
constant and the mass square value is well approximated by
$M^2(T<T_c)=\hat{\Lambda}^2/{2\sqrt{e}c}$ .

\begin{figure}[hbtp]
\begin{center}
\includegraphics[height=6cm,width=10cm, angle=0]{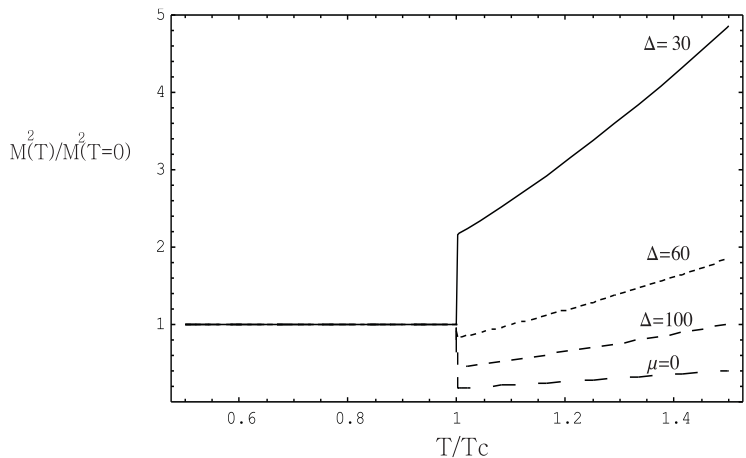}
\end {center}
\caption{Illustrative plot of the glueball mass as function of the
temperature , $M^2(T)/M^2(T=0)$ for different values of $\Delta$
and fixed chemical potential $\mu=500$~MeV. The curve labelled by
$\mu=0$ corresponds to the in vacuum (but hot) case. It is
straightforward to consider another value of $\mu$ and the
qualitative picture remains unchanged.} \label{figure3}
\end{figure}

Although the glueball treatment alone cannot be used above the
deconfining phase transition it is nevertheless interesting to
consider such a temperature region. For $T\geq T_c$ the new global
minimum is at zero and we can use Eq.~(\ref{exactTP}) to deduce
\begin{eqnarray} \frac{M^2(T\geq T_c)}{M^2(T=0)}=
\frac{\sqrt{5}}{4\pi \sqrt{v^3}}\left[\frac{T}{T_c}\right]^2 \
.\label{squaremass}
\end{eqnarray}
Due to the velocity factor in Eq.~(\ref{squaremass}) there is a
relative enhancement with to the respect to the in vacuum (but
hot) Yang-Mills theory. {}For illustration  we plot our results in
Fig.~\ref{figure3} for the in vacuum (i.e. $v=1$ and $\mu=0$) and
the in medium theory for $\mu=500$ while considering different
values of $\Delta$.
Since the glueball mass increases in the deconfined phase the new
light degrees of freedom (namely the elementary gluons themselves)
now dominate the free energy. Interestingly we find that due to a
large dielectric constant of the 2SC medium the associated light
glueballs mass square in the deconfined region gains a factor
$1/\sqrt{v^3}$ relative to the in vacuum case. Hence for all of
the relevant thermodynamical properties/quantities of the 2SC
above the deconfining $SU(2)$ phase transition the glueballs are
not expected to play an important role. Since we plot the ratio of
masses the result does not depend on the positive constant $c$.
{}From the figure is also clear that there is a strong dependence
on the specific value of $\Delta$.

\begin{figure}[hbtp]
\begin{center}
\includegraphics[angle=0]{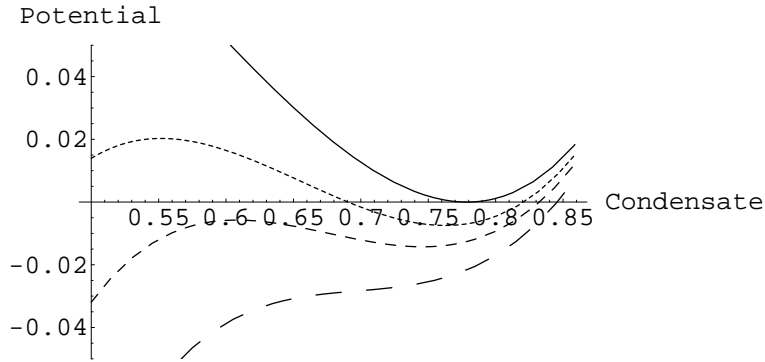}
\end {center}
\caption{Zoom of the Potential of Eq.~(\ref{thermalP}) close to
the old minimum for $\mu=500$ MeV, $\Delta=30$ MeV as function of
the condensate $\bar{h}/\hat{\Lambda}$ for different values of the
temperature above $T_c$. The solid line corresponds to $T=T_c$,
the dotted line to $T=1.8~T_c$, the short-dashed to $T=1.9~T_c$
and the long dashed to $T=2~T_c$. We have, as in
Fig.~\ref{figure1}, chosen $c=1/(50 \sqrt{e})$.} \label{figure4}
\end{figure}

We now comment briefly on the fate of the old minimum as the
temperature is increased above the critical temperature and in
absence of elementary gluons. The value of the glueball condensate
corresponding to the old minimum just above the critical
temperature start reducing while the minimum disappears for a
temperature of the order of $\approx 2T_c$. This behavior is
summarized in Fig.~\ref{figure4} and it is a classical example of
first order phase transition if we were to consider the glueball
Lagrangian as the correct description at and above the transition
point.
When restricting to the in vacuum theory our results are in
reasonable agreement with the results and expectations of various
investigations (see for example \cite{Carter:1998ti} and
references therein) using similar models.

The glueball Lagrangian based model cannot be used to establish
the order of the phase transition since the gluon condensate is
not an order parameter for a Yang-Mills theory although it does
encode information on the underlying conformal anomaly. The break
down point signals the presence of new lighter degrees of freedom
which needed to be taken into account. In the absence of quarks a
reasonable order parameter for the $SU(N)$ Yang-Mills theory is
the Polyakov loop \cite{Svetitsky:1982gs}:
\begin{eqnarray} {\ell}\left(x\right)=\frac{1}{N}{\rm Tr}({\bf L})\equiv\frac{1}{N}{\rm Tr}
\left[{\cal
P}\exp\left[i\,g\int_{0}^{1/T}A_{0}(x,\tau)d\tau\right]\right] \ ,
\end{eqnarray} where ${\cal P}$ denotes path ordering, $g$ is the $SU(N)$ coupling constant and $x$ is
the coordinate for the three spatial dimensions while $\tau$ is
euclidean time. The $\ell$ field is real for $N=2$ while otherwise
complex.  This object is charged with respect to the center $Z_N$
of the $SU(N)$ gauge group \cite{Svetitsky:1982gs} under which it
transforms as $\ell \rightarrow z \ell$ with $z\in Z_N$. A
relevant feature of the Polyakov loop is that its expectation
vanishes in the low temperature regime and is non zero in the high
temperature phase.

This behavior has recently lead Pisarski \cite{Pisarski:2001pe} to
model the Yang-Mills (non supersymmetric) phase transition as a
mean field theory of Polyakov loops. One can simply show that for
$SU(2)$ one expects a second order phase transition (as function
of the temperature) and a weak first order for $SU(3)$.

We can use Pisarski's model to predict the order of the transition
in the present case. Assuming that a local $SU(2)$ Yang-Mills
action at low energies does exists we construct the simplest
Polykov loop using the rescaled space time coordinates and fields
\cite{{rischke2k},OS2}:
\begin{eqnarray} {\hat{\ell}}\left(x\right)=\frac{1}{2}{\rm Tr}
\left[{\cal
P}\exp\left[i\,\hat{g}\int_{0}^{1/\hat{T}}\hat{A}_{0}({x},\hat{\tau})d\hat{\tau}\right]\right]
\ ,
\end{eqnarray}
with $\hat{A_0}=\hat{A}_0^a \tau^a/2$ and $\tau^a$ the $SU(2)$
Pauli matrices, and the connection with the underlying fields is
\cite{{rischke2k},OS2}
$\hat{A_0^a}=\lambda^{\frac{1}{4}}\epsilon^{\frac{3}{4}}A_0^a$
while $\hat{g}=g_{s}(\lambda/\epsilon)^{\frac{1}{4}}$. The
rescaled euclidean time $\hat{\tau}=\tau/\sqrt{\lambda \epsilon}$
leads to $\hat{T}=T/v$ while $\lambda$ is a possible magnetic
permeability which turns to be equal to one in our case. At this
point the effective mean field type of model a l\'a Pisarski for
$\hat{\ell}$ is similar to the one for the in vacuum $SU(2)$
Yang-Mills theory. So if we make the strong but plausible
assumption (as shown above) that all the way up and above the
deconfinement $SU(2)$ color phase transition the effects of the
quark superconductive matter can be taken into account just via a
non zero dielectric constant we expect a second order phase
transition for the $SU(2)$ of color in 2SC. It is relevant to
mention that the order of the transition might change if we
include new contributions arising for example by considering the
quark masses.  The deconfining temperature is expected to be the
one close to our prediction obtained from the glueball model
Lagrangian. Even if a non zero magnetic permeability exists the
present argument would not be modified.

$SU(2)$ Yang-Mills with non zero dielectric and magnetic
permeability can be simulated, using standard sampling methods, on
the lattice. For a large body of work on $SU(2)$--Yang--Mills
theory we refer to \cite{Damgaard}. These results would test at
the same time the validity of the glueball model for the
prediction of the critical temperature and the order of the phase
transition according to the Polyakov loop model in a framework
slightly modified with respect to the in vacuum case. Besides the
latter would also constitute the first lattice simulations testing
the high quark chemical potential but small temperature region of
the QCD phase diagram.

The disagreement between the first order phase transition
predicted by the glueball Lagrangian and the previous argument
based on the symmetries obeyed by the order parameter is an
apparent one. In fact any gauge invariant quantity which is not an
order parameter does not need to behave as the order parameter
itself at the transition \cite{Pisarski:2001pe} as discussed at
length in the introduction (see also \cite{Sannino:2002wb}).

\section{Conclusions}
\label{Conclusions} We studied the temperature effects on the
unbroken $SU(2)$ color gauge interactions for the 2 flavor case at
high matter density. Using a simple model based on a light
glueball Lagrangian we estimated the $SU(2)$ deconfinement
critical temperature for given chemical potential and
superconductive gap value. We have shown that the
deconfining/confining critical temperature is smaller than the
critical temperature for the superconductive state itself. The
breaking of Lorentz invariance (already at zero temperature),
encoded in the glueball velocity, further reduces the critical
temperature by a factor $v^{3/4}$ relative to the in vacuum case.
By computing the glueball thermal effective potential we have the
equation of state for part of the ideal 2SC phase (i.e. zero up
and down quark masses and infinitely massive strange quark). In
particular we can compute the pressure, the energy density and the
entropy of the system.

Another relevant point is that we have developed a general
framework according to which any parameterization of the glueball
field can be used to construct the full thermal effective action
arising from the lagrangian constructed using the anomalous
variation of dilation current.

Using the Polyakov loop model, adapted to the present case we also
predict the associated phase transition to be second order.

In order to apply our model to the physics of compact objects we
should extend it in order to take into accounts the  effects of
the the up and down quark masses as well as the effects of a not
too massive strange quark.

\subsection*{Notes added in proof}

About two months after we submitted this paper the paper
\cite{Alford:2002kj} appeared where it is claimed that the 2SC
state might not be present on compact stars. This is a very
dynamical issue which deserves further studies. The present paper
deals with the properties of part of the ideal 2SC state and as
such our results are not affected by this claim. However possible
astrophysical applications may be affected. Finally also
Ref.~\cite{Sannino:2002wb} which clarifies the relation between
the order parameter and the gluon condensate further strengthening
our approach appeared after this paper was submitted.

 \acknowledgments It is a pleasure for us to thank R.
Casalbuoni for suggesting this problem to us. We would like to
thank P. Damgaard for enlightening discussions, J. Schechter for
discussions and reading of the manuscript, and K. Splittorff for
interesting discussions. We also acknowledge discussions with C.
Manuel,  R. Ouyed and O. Scavenius. The work of F.S. is supported
by the Marie--Curie fellowship under contract MCFI-2001-00181,
N.M. was supported by the EU Comission under contract
HPMT-2000-00010 and by NORDITA, while W.S. acknowledges support by
DAAD and NORDITA.

\end{document}